\documentclass[12pt]{article}

\usepackage{graphics}
\usepackage{epsfig}
\input epsf

\title{Prompt photon hadroproduction at high energies \\ in off-shell gluon-gluon fusion}

\author{S.P.~Baranov$^a$, A.V.~Lipatov$^b$, N.P.~Zotov$^b$}

\begin{document}

\maketitle

\begin{center}

{\it $^a$\,P.N.~Lebedev Physics Institute,\\ 
119991 Moscow, Russia\/}\\[3mm]

{\it $^b$\,D.V.~Skobeltsyn Institute of Nuclear Physics,\\ 
M.V. Lomonosov Moscow State University,
\\119991 Moscow, Russia\/}\\[3mm]

\end{center}

\vspace{0.5cm}

\begin{center}

{\bf Abstract }

\end{center}

The amplitude for production of a single photon associated with 
quark pair in the fusion of two off-shell gluons is calculated. 
The matrix element found is applied to the inclusive 
prompt photon hadroproduction 
at high energies in the framework of $k_T$-factorization QCD
approach. The total and differential cross sections are
calculated in both central and forward pseudo-rapidity regions.
The conservative error analisys is performed.
We used the unintegrated gluon distributions in a proton which were
obtained from the full CCFM evolution equation as well as from the
Kimber-Martin-Ryskin prescription.
Theoretical results were compared with recent
experimental data taken by the D$\oslash$ and CDF collaborations at 
Fermilab Tevatron. Theoretical predictions for the LHC energies are 
given.

\vspace{1cm}

\noindent
PACS number(s): 12.38.-t, 13.85.-t

\vspace{0.5cm}

\section{Introduction} \indent 

The production of prompt (or direct) photons in hadron-hadron collisions at the
Tevatron is a subject of intens studies~[1--6]. Usually the photons are 
called "prompt" if they are coupled to the interacting quarks.
The theoretical and experimental investigations of such processes have 
provided a direct probe of the hard subprocess dynamics, since the produced 
photons are largely insensitive to the effects of final-state hadronization.
At the leading order, prompt photons  
can be produced via quark-gluon Compton scattering or quark-antiquark 
annihilation~[7] and so, the cross sections of these processes are 
strongly sensitive to the parton (quark and gluon) content of a 
proton\footnote{Also observed photon may arise 
from so called fragmentation processes~[8], where a final state quark or 
gluon decays into $\gamma$. This contribution will be discussed below in Section~2.}.
Very recently experimental data~[6] on the inclusive prompt photon hadroproduction
at the Tevatron have been presented 
by the D$\oslash$ collaboration. These data extend previous measurements
to significantly higher values of the photon $p_T$ 
(namely to $p_T \sim 300$~GeV at $\sqrt s = 1960$~GeV).
In the present paper we will analyse the data~[1--6]
using the so-called $k_T$-factorization~[9, 10] (or semihard~[11, 12])
approach of QCD.

The $k_T$-factorization approach is based on the familiar 
Balitsky-Fadin-Kuraev-Lipatov (BFKL)~[13] or Ciafaloni-Catani-Fiorani-Marchesini 
(CCFM)~[14] gluon evolution equations. In this way, the large logarithmic terms 
proportional to $\ln 1/x$ are summed up to all orders of perturbation theory 
(in the leading logarithm approximation). It is in contrast with the popular 
Dokshitzer-Gribov-Lipatov-Altarelli-Parizi (DGLAP)~[15] strategy where only large 
logarithmic terms proportional to $\ln \mu^2$ are taken into account. 
The basic dynamical quantity of the $k_T$-factorization approach is 
the so-called unintegrated (i.e. ${\mathbf k}_T$-dependent) gluon distribution 
${\cal A}(x,{\mathbf k}_T^2,\mu^2)$ which determines the probability to find a 
gluon carrying the longitudinal momentum fraction $x$ and the transverse momentum 
${\mathbf k}_T$ at the probing scale $\mu^2$. The unintegrated gluon distribution
can be obtained from the analytical or numerical solution of BFKL or CCFM
equations. Similar to DGLAP, to calculate the cross sections of any 
physical process the unintegrated gluon density ${\cal A}(x,{\mathbf k}_T^2,\mu^2)$ 
has to be convoluted~[9--12] with the relevant partonic cross section 
which has to be taken off mass shell (${\mathbf k}_T$-dependent). 
It is in clear contrast with the usual DGLAP scheme (so-called collinear factorization). 
Since gluons in the initial state are not on-shell and are characterized by virtual 
masses (proportional to their transverse momentum), it also assumes a modification 
of their polarization density matrix~[11, 12]. In particular, the polarization 
vector of a gluon is no longer purely transversal, but acquires an admixture of 
longitudinal and time-like components. Other important properties of the 
$k_T$-factorization formalism are the additional contribution to the cross 
sections due to the integration over the ${\mathbf k}_T^2$ region above $\mu^2$
and the broadening of the transverse momentum distributions due to extra 
transverse momentum of the colliding partons.

It is important that at present a complete theoretical description 
of the prompt photon production data at the Tevatron is a subject of special 
investigations (see, for example,~[16] and references therein).
Both the completeness of next-to-leading order (NLO) 
perturbative QCD calculations~[17] and consistency of the available 
data have been the 
subject of intense discussions~[18--25]. Despite that 
the NLO pQCD predictions agree with the recent high-$p_T$ measurements~[6] within 
uncertainties, there are still open questions.
So, it was found~[1--4] that the shape of the measured cross sections as a 
function of photon transverse energy $E_T$ is poorly described by the NLO pQCD 
calculations: the observed $E_T$ distribution is steeper than the 
predictions of perturbative QCD. These shape differences lead to a significant 
disagreement in the ratio of cross sections calculated at different center-of-mass
energies $\sqrt s = 630$ GeV and $\sqrt s = 1800$ GeV as a function of scaling 
variable $x_T = 2 E_T^\gamma/\sqrt s$. It was claimed~[2, 3] that the disagreement 
in the $x_T$ ratio is difficult to explain with conventional theoretical 
uncertainties connected with the scale dependence and parametrizations of the parton 
distributions. The origin of the disagreement has been attributed to the effect of 
initial-state soft-gluon radiation~[21, 22]. In the papers~[22, 26] it was shown that 
the observed discrepancy 
can be reduced by introducing some additional intrinsic transverse momentum $k_T$ 
of the incoming partons, which is usually assumed to have a Gaussian-like 
distribution. The average value of this $k_T$ increases from 
$\langle k_T \rangle \sim 1$ GeV to more than 
$\langle k_T \rangle \sim 3$ GeV in  hard-scattering processes as 
the $\sqrt s$ increases from UA6 to Tevatron energies~[22, 25].
The importance of including the gluon emission through the resummation 
formalism was recognized and only recently this approach has been 
developed for inclusive prompt photon production~[26--30].

In the framework of $k_T$-factorization formalism the treatment 
of $k_T$-enhancement in the inclusive prompt photon hadroproduction 
at Tevatron suggests a modification of the above simple $k_T$ 
smearing picture. In this approach the transverse momentum of incoming partons
is generated in the course of non-collinear parton evolution under control of 
relevant evolution equations.
In paper~[18] the Kimber-Martin-Ryskin (KMR) formalism~[31] (in the double 
leading logarithmic approximation) was applied to study the role of the both non-perturbative
and perturbative components of partonic $k_T$ in describing of the observed 
$E_T$ spectrum. Note that this formalism
is based on the standard DGLAP evolution. 
In calculations~[18] the usual on-shell matrix elements 
of quark-gluon fusion and quark-antiquark annihilation were evaluated with 
precise off-shell kinematics. In our previous paper~[32] 
we have used the proper off-shell 
expressions for the matrix elements of these partonic subprocesses
and also the KMR-constructed unintegrated parton densities\footnote{A similar
scenario has been used also in~[33].} (which were evaluated 
independently from other authors).
Our predictions for the inclusive prompt photon production 
agree well with available D$\oslash$~[1, 2] and CDF~[3--5] experimental data at Tevatron 
in both central and forward 
pseudo-rapidity regions. Perfect agreement was found also in the ratio of two cross 
sections calculated at $\sqrt s = 630$ GeV and $\sqrt s = 1800$ GeV. 
This ratio shows specific effect connected with off-shell gluons in the
$k_T$-factorization approach.

We note, however, that an important component of all above calculations~[18, 32] is the 
unintegrated quark distribution in a proton. At present these densities
are available in the framework of KMR approach only. 
It is because there are some theoretical difficulties 
in obtaining the quark distributions immediately from CCFM or BFKL 
equations\footnote{Unintegrated quark density was considered recently in~[34].}
(see, for example, review~[35] for more details).
As a result the dependence of the calculated cross sections 
on the non-collinear evolution scheme has not been investigated yet.
This dependence in general can be significant and it is a special 
subject of study in the $k_T$-factorization formalism. 
Therefore in the present paper we will try a different and more systematic way.
Instead of using the unintegrated quark distributions and 
the corresponding quark-gluon fusion and quark-antiquark annihilation cross
sections we will calculate off-shell matrix element of 
the $g^* g^* \to q \bar q \gamma$ subprocess
and then will operate in terms of the unintegrated gluon 
densities only. In this way the different non-collinear 
evolution schemes can be applied. 
But we note that this matrix element covers only sea quark contribution 
from the quark-gluon fusion and quark-antiquark annihilation and
therefore in this case contribution from the valence quarks is not taken into account.
However, this contribution is significant only at large $x$
and therefore can be safely accounted for in the collinear LO approximation as 
additional one.

So, in the present paper we will investigate the different prompt 
photon production rates using the off-mass shell matrix elements 
of the $g^* g^* \to q \bar q \gamma$ subprocess 
and make a systematic comparison of our predictions to the available
D$\oslash$ and CDF data~[1--6]. Our special goal is to
study the sensitivity of the calculated cross sections
to the non-collinear evolution scheme.
In the numerical calculations we will test the different sets of 
unintegrated gluon distributions which are 
obtained from the full CCFM equation as well as from the
Kimber-Martin-Ryskin approach. Additionally we give
some predictions for LHC conditions.

The outline of our paper is following. In Section~2 we 
recall shortly the basic formulas of the $k_T$-factorization approach with a brief 
review of calculation steps. In Section~3 we present the numerical results
of our calculations and a discussion. Finally, in Section~4, we give
some conclusions.

\section{Theoretical framework} 
\subsection{Kinematics} \indent 

We start from the kinematics (see Fig.~1). 
Let $p^{(1)}$ and $p^{(2)}$ be the four-momenta of the incoming protons and 
$p$ be the four-momentum of the produced photon.
The initial off-shell gluons have the four-momenta
$k_1$ and $k_2$ and the final quark and antiquark have the 
four-momenta $p_1$ and $p_2$ and masses $m_q$, respectively.
In the $p \bar p$ center-of-mass frame we can write
$$
  p^{(1)} = {\sqrt s}/2\,(1,0,0,1),\quad p^{(2)} = {\sqrt s}/2\,(1,0,0,-1), \eqno(1)
$$

\noindent
where $\sqrt s$ is the total energy of the process 
under consideration and we neglect the masses of the incoming protons.
The initial gluon four-momenta in high energy limit can be written as
$$
  k_1 = x_1 p^{(1)} + k_{1T},\quad k_2 = x_2 p^{(2)} + k_{2T}, \eqno(2)
$$

\noindent 
where $k_{1T}$ and $k_{2T}$ are their transverse four-momenta.
It is important that ${\mathbf k}_{1T}^2 = - k_{1T}^2 \neq 0$ and
${\mathbf k}_{2T}^2 = - k_{2T}^2 \neq 0$. From the conservation laws 
we can easily obtain the following conditions:
$$
  {\mathbf k}_{1T} + {\mathbf k}_{2T} = {\mathbf p}_{1T} + {\mathbf p}_{2T} + {\mathbf p}_{T},
$$
$$
  x_1 \sqrt s = m_{1T} e^{y_1} + m_{2T} e^{y_2} + |{\mathbf p}_T| e^y, \eqno(3)
$$
$$
  x_2 \sqrt s = m_{1T} e^{-y_1} + m_{2T} e^{-y_2} + |{\mathbf p}_T| e^{-y},
$$

\noindent 
where $y$ is the rapidity of produced photon, 
$p_{1T}$ and $p_{2T}$ are the transverse four-momenta of final quark and antiquark, 
$y_1$, $y_2$, $m_{1T}$ and $m_{2T}$ are their center-of-mass rapidities and 
transverse masses, i.e. $m_{iT}^2 = m_q^2 + {\mathbf p}_{iT}^2$.

\subsection{Off-shell amplitude of the $g^* g^* \to q \bar q \gamma$ subprocess} \indent 

There are eight Feynman diagrams (see Fig.~2) which describe the partonic
subprocess $g^* g^* \to q \bar q \gamma$ at the leading order in $\alpha_s$ and $\alpha$.
Let $\epsilon_1$, $\epsilon_2$ and $\epsilon$ be the initial gluon and produced photon 
polarization vectors, respectively, and $a$ and $b$ the eight-fold color indices of the off-shell
gluons.
Then the relevant matrix element can be presented as follows:
$$
  {\cal M}_1 = e g^2 \, \bar u (p_1) \, t^a \gamma^\mu \epsilon_\mu {\hat p_1 - \hat k_1 + m_q\over m_q^2 - (p_1 - k_1)^2} \gamma^\lambda \epsilon_\lambda {\hat k_2 - \hat p_2 + m_q\over m_q^2 - (k_2 - p_2)^2} t^b \gamma^\nu \epsilon_\nu \, u(p_2), \eqno(4)
$$
$$
  {\cal M}_2 = e g^2 \, \bar u (p_1) \, t^b \gamma^\nu \epsilon_\nu {\hat p_1 - \hat k_2 + m_q\over m_q^2 - (p_1 - k_2)^2} \gamma^\lambda \epsilon_\lambda {\hat k_1 - \hat p_2 + m_q\over m_q^2 - (k_1 - p_2)^2} t^a \gamma^\mu \epsilon_\mu \, u(p_2), \eqno(5)
$$
$$
  {\cal M}_3 = e g^2 \, \bar u (p_1) \, t^a \gamma^\mu \epsilon_\mu {\hat p_1 - \hat k_1 + m_q\over m_q^2 - (p_1 - k_1)^2}\, t^b \gamma^\nu \epsilon_\nu { - \hat p_2 - \hat p + m_q\over m_q^2 - ( - p_2 - p)^2} \gamma^\lambda \epsilon_\lambda \, u(p_2), \eqno(6)
$$
$$
  {\cal M}_4 = e g^2 \, \bar u (p_1) \, t^b \gamma^\nu \epsilon_\nu {\hat p_1 - \hat k_2 + m_q\over m_q^2 - (p_1 - k_2)^2}\, t^a \gamma^\mu \epsilon_\mu { - \hat p_2 - \hat p + m_q\over m_q^2 - ( - p_2 - p)^2} \gamma^\lambda \epsilon_\lambda \, u(p_2), \eqno(7)
$$
$$
  {\cal M}_5 = e g^2 \, \bar u (p_1) \, \gamma^\lambda \epsilon_\lambda {\hat p_1 + \hat p + m_q\over m_q^2 - (p_1 + p)^2}\, t^b \gamma^\nu \epsilon_\nu { \hat k_1 - \hat p_2 + m_q\over m_q^2 - (k_1 - p_2)^2} t^a \gamma^\mu \epsilon_\mu \, u(p_2), \eqno(8)
$$
$$
  {\cal M}_6 = e g^2 \, \bar u (p_1) \, \gamma^\lambda \epsilon_\lambda {\hat p_1 + \hat p + m_q\over m_q^2 - (p_1 + p)^2}\, t^a \gamma^\mu \epsilon_\mu { \hat k_2 - \hat p_2 + m_q\over m_q^2 - (k_2 - p_2)^2} t^b \gamma^\nu \epsilon_\nu \, u(p_2), \eqno(9)
$$
$$
  \displaystyle {\cal M}_7 = e g^2 \, \bar u (p_1) \, \gamma^\rho C^{\mu \nu \rho}(k_1,k_2,- k_1 - k_2){\epsilon_\mu \epsilon_\nu \over (k_1 + k_2)^2} f^{abc} t^c \times \atop 
  \displaystyle \times { - \hat p_2 - \hat p + m_q\over m_q^2 - ( - p_2 - p)^2}\, \gamma^\lambda \epsilon_\lambda \, u(p_2), \eqno(10)
$$
$$
  \displaystyle {\cal M}_8 = e g^2 \, \bar u (p_1) \, \gamma^\lambda \epsilon_\lambda {\hat p_1 + \hat p + m_q\over m_q^2 - (p_1 + p)^2} \times \atop 
  \displaystyle \times \gamma^\rho C^{\mu \nu \rho}(k_1,k_2,- k_1 - k_2) {\epsilon_\mu \epsilon_\nu \over (k_1 + k_2)^2} f^{abc} t^c \, u(p_2). \eqno(11)
$$

\vspace{0.2cm}

\noindent
In above expressions $C^{\mu \nu \rho}(k,p,q)$ is related to the standard QCD
three-gluon coupling 
$$
  C^{\mu \nu \rho}(k,p,q) = g^{\mu \nu} (p - k)^\rho + g^{\nu \rho} (q - p)^\mu + g^{\rho \mu} (k - q)^\nu. \eqno(12)
$$

\noindent
The summation on the produced photon polarization is carried out by
covariant formula
$$
  \sum \epsilon^\mu \epsilon^{* \, \nu} = - g^{\mu \nu}. \eqno(13)
$$

\noindent
In the case of initial off-shell gluon we use the BFKL prescription~[11--13]:
$$
  \epsilon^\mu (k) \epsilon^{* \, \nu} (k) = {k_T^\mu k_T^\nu \over {\mathbf k}_T^2}. \eqno(14)
$$

\noindent
This formula converges to the usual one~(13) after azimuthal angle averaging
in the $k_T \to 0$ limit. 
The evaluation of the traces in~(4) --- (11) was done using the algebraic 
manipulation system FORM~[36]. 
The usual method of squaring of (4) --- (11) results in enormously long
output. This technical problem was solved by applying the
method of orthogonal amplitudes~[37].

The gauge invariance of matrix element is a
subject of special attension in the $k_T$-factorization approach. Strictly speaking,
the diagrams shown in Fig.~2 are insufficient and have to be accompanied
with the graphs involving direct gluon exchange between the protons
(these protons are not shown in Fig.~2). These graphs are 
necessary to maintain the gauge invariance.
However, they violate the factorization since they cannot be represented
as a convolution of gluon-gluon fusion matrix element with unintegrated gluon density.
The solution pointed out in~[10] refers to the fact that, within the 
particular gauge (13), the contribution from these unfactorizable diagrams
vanish, and one has to only take into account the graphs depicted in Fig.~2.
We have succesfully tested the gauge invariance of matrix element~(4) --- (11) numerically.

\subsection{Fragmentation contributions} \indent 

Perturbation theory becomes nonapplicable when the wavelength of the 
emitted photon (in the emitting quark rest frame) becomes larger that the 
typical hadronic scale ${\cal{O}}$ (1 GeV$^{-1}$). Then the nonperturbative 
effects of hadronization or fragmentation must be taken into account. 
Accordingly, the calculated cross section can be split into two pieses 
$$
  d\sigma = d\sigma_{\mbox{direct}}(\mu^2) + d\sigma_{\mbox{fragm}}(\mu^2)
$$ 

\noindent
with $d\sigma_{\mbox{direct}}(\mu^2)$ representing the perturbative 
contribution and $d\sigma_{\mbox{fragm}}(\mu^2)$ the fragmentation 
contribution. In our calculations we choose the fragmentationn scale 
$\mu^2$ to be the invariant mass of the quark + photon subsystem, 
$\mu^2 = (p + p_i)^2$, and restrict $d\sigma_{\mbox{direct}}(\mu^2)$ 
to $\mu\ge M\simeq 1$~GeV. Under this condition, the contribution 
$d\sigma_{\mbox{direct}}(\mu^2)$ is free from divergences, so that 
the mass of the light quark $m_q$ can be safely sent to zero. The sensitivity 
of our results to the choice of $M$ is reasonably soft, as we will discuss 
in Section~3. As far as the fragmentation contribution is concerned, its 
size is dramatically reduced by the photon isolation cuts (see below). 
According to the estimates presented in Ref.~[38], the contribution 
from $d\sigma_{\mbox{fragm}}$ amounts to about 10\% of the visible cross 
section. This value is smaller than the theoretical uncertainty in 
calculating the perturbative contribution $d\sigma_{\mbox{direct}}$, and so, is 
neglected in our analysis. 

\subsection{Photon isolation cuts} \indent 

In order to reduce huge background
from the secondary photons produced by the decays of $\pi^0$ and $\eta$ 
mesons the isolation criterion is introduced in the experimental analyses.
This criterion is the following. A photon is isolated if the 
amount of hadronic transverse energy $E_T^{\rm had}$, deposited inside
a cone with aperture $R$ centered around the photon direction in the 
pseudo-rapidity and azimuthal angle plane, is smaller than
some value $E_T^{\rm max}$:
$$
  \displaystyle E_T^{\rm had} \le E_T^{\rm max},\atop
  \displaystyle (\eta^{\rm had} - \eta)^2 + (\phi^{\rm had} - \phi)^2 \le R^2. \eqno(15)
$$

\noindent 
The both D$\oslash$ and CDF collaborations take $R \sim 0.4$ and 
$E_T^{\rm max} \sim 1$ GeV in the experiment~[1--6]. 
Isolation not only reduces the background 
but also significantly reduces the fragmentation components
connected with collinear photon radiation. 
It was shown that after applying the isolation cut~(15) the 
contribution from the fragmentation subprocesses is strongly suppressed~[38].

\subsection{Cross section for prompt photon hadroproduction} \indent 

In general case, according to $k_T$-factorization theorem the 
cross section of prompt photon hadroproduction 
can be written as a convolution
$$
  \displaystyle \sigma (p \bar p \to \gamma\,X) = \sum_{q} \int {dx_1\over x_1} {\cal A}(x_1,{\mathbf k}_{1 T}^2,\mu^2) d{\mathbf k}_{1 T}^2 {d\phi_1\over 2\pi} \times \atop 
  \displaystyle \times \int {dx_2\over x_2} {\cal A}(x_2,{\mathbf k}_{2 T}^2,\mu^2) d{\mathbf k}_{2 T}^2 {d\phi_2\over 2\pi} d{\hat \sigma} (g^* g^* \to q \bar q \gamma), \eqno(16)
$$

\noindent 
where $\hat \sigma(g^* g^* \to q \bar q \gamma)$ is the partonic cross section, 
${\cal A}(x,{\mathbf k}_{T}^2,\mu^2)$ is the unintegrated gluon distribution in a proton 
and $\phi_1$ and $\phi_2$ are the azimuthal angles of the incoming gluons.
The multiparticle phase space $\Pi d^3 p_i / 2 E_i \delta^{(4)} (\sum p^{\rm in} - \sum p^{\rm out} )$
is parametrized in terms of transverse momenta, rapidities and azimuthal angles:
$$
  { d^3 p_i \over 2 E_i} = {\pi \over 2} \, d {\mathbf p}_{iT}^2 \, dy_i \, { d \phi_i \over 2 \pi}. \eqno(17)
$$

\noindent
Using the expressions~(16) and~(17) we can easily obtain the master formula:
$$
  \displaystyle \sigma(p \bar p \to \gamma\,X) = \sum_{q} \int {1\over 256\pi^3 (x_1 x_2 s)^2} |\bar {\cal M}(g^* g^* \to q \bar q\gamma)|^2 \times \atop 
  \displaystyle \times {\cal A}(x_1,{\mathbf k}_{1 T}^2,\mu^2) {\cal A}(x_2,{\mathbf k}_{2 T}^2,\mu^2) d{\mathbf k}_{1 T}^2 d{\mathbf k}_{2 T}^2 d{\mathbf p}_{1 T}^2 {\mathbf p}_{2 T}^2 dy dy_1 dy_2 {d\phi_1\over 2\pi} {d\phi_2\over 2\pi} {d\psi_1\over 2\pi} {d\psi_2\over 2\pi}, \eqno(18)
$$

\noindent
where $|\bar {\cal M}(g^* g^* \to q\bar q \gamma)|^2$ is the off-mass shell 
matrix element squared and averaged over initial gluon 
polarizations and colors, $\psi_1$ and $\psi_2$ are the 
azimuthal angles of the final state quark and antiquark, respectively.
We would like to point out again that $|\bar {\cal M}(g^* g^* \to q\bar q \gamma)|^2$
is strongly depends on the non-zero 
transverse momenta ${\mathbf k}_{1 T}^2$ and ${\mathbf k}_{2 T}^2$.
If we average the expression~(18) over $\phi_{1}$ and $\phi_{2}$ 
and take the limit ${\mathbf k}_{1 T}^2 \to 0$ and ${\mathbf k}_{2 T}^2 \to 0$,
then we recover the expression for the prompt photon hadroproduction in the usual 
collinear approximation.

The multidimensional integration in~(18) has been performed
by the means of Monte Carlo technique, using the routine 
\textsc{Vegas}~[39]. The full C$++$ code is available from the 
authors on request\footnote{lipatov@theory.sinp.msu.ru}.

\section{Numerical results}
\subsection{Theoretical uncertainties} \indent 

Except the the unintegrated gluon distribution in a 
proton ${\cal A}(x,{\mathbf k}_T^2,\mu^2)$,
there are several parameters which determined the overall 
normalization factor of the cross section~(18): the quark 
mass $m_q$, the factorization and normalisation scales 
$\mu_F$ and $\mu_R$.

Concerning the unintegrated gluon densities in a proton, 
we have tried here two different sets of them. These 
sets are widely discussed in the literature 
(see, for example, review~[35] for more information). 
Here we only shortly discuss their characteristic properties.

First unintegrated gluon density used has been obtained~[40] recently
from the numerical solution of the full CCFM equation. 
Function ${\cal A}(x,{\mathbf k}_T^2,\mu^2)$ is determined
by a convolution of the non-perturbative starting
distribution ${\cal A}_0(x)$ and the CCFM evolution
denoted by $\tilde {\cal A}(x,{\mathbf k}_T^2,\mu^2)$:
$$
  {\cal A}(x,{\mathbf k}_T^2,\mu^2) = \int {d x'\over x'} {\cal A}_0(x') \tilde {\cal A}\left({x\over x'},{\mathbf k}_T^2,\mu^2\right). \eqno(19)
$$

\noindent
In the perturbative evolution the gluon splitting function
$P_{gg}(z)$ including non-singular terms (as described in detail in~[41])
is applied. The input parameters in ${\cal A}_0(x)$
were fitted to describe the proton structure function $F_2(x,Q^2)$.
A acceptable fit to the measured $F_2$ values was obtained with
$\chi^2/ndf = 1.83$ using statistical and uncorrelated systematical
uncertainties (compare to $\chi^2/ndf \sim 1.5$ in the collinear approach
at NLO). This distribution has been applied recently in the analysis of the 
deep inelastic proton structure functions $F_2^c$, $F_2^b$ and
$F_L$~[42].

Another set (the so-called KMR distribution) 
is the one which was originally proposed in~[31]. The KMR approach 
is the formalism to construct unintegrated gluon distribution from the 
known conventional parton (quark and gluon) densities. 
It accounts for the angular-ordering (which comes from 
the coherence effects in gluon emission) as well as the main part of the 
collinear higher-order QCD corrections. The key observation here 
is that the $\mu$ dependence of the unintegrated parton distribution 
enters at the last step of the evolution. In 
the numerical calculations we have used the 
standard GRV~(LO) parametrizations~[43] of the collinear quark and gluon 
densities. The KMR-constructed parton distributions were used, 
in particular, to describe the prompt photon photoproduction at 
HERA~[44] and prompt photon hadroproduction Tevatron~[32].

Significant theoretical uncertainties are connected with the
choice of the factorization and renormalization scales. The first of them
is related to the evolution of the gluon distributions, the other is 
responsible for the strong coupling constant $\alpha_s(\mu^2_R)$.
As it often done, we choose the 
renormalization and factorization scales to be equal: 
$\mu_R = \mu_F = \mu = \xi |{\mathbf p}_{T}|$.
In order to investigate the scale dependence of our 
results we will vary the scale parameter
$\xi$ between $1/2$ and 2 about the default value $\xi = 1$.

The masses of all light quarks were set to be equal to 
$m_q = 4.5$~MeV and the charmed quark mass was set to 
$m_c = 1.4$~GeV. We have checked that uncertainties which come 
from these quantities are negligible in comparison to the uncertainties
connected with the unintegrated gluon distributions.
For completeness, we use LO formula for the strong 
coupling constant $\alpha_s(\mu^2)$ with $n_f = 3$ 
active quark flavours at $\Lambda_{\rm QCD} = 232$~MeV, 
such that $\alpha_s(M_Z^2) = 0.117$.
Note that we use special choice $n_f = 4$ and $\Lambda_{\rm QCD} = 130$~MeV 
in the case of CCFM gluon ($\alpha_s(M_Z^2) = 0.118$), 
as it was originally proposed in~[40]. 

\subsection{Inclusive prompt photon production at Tevatron} \indent

The experimental data~[1--6] for the inclusive prompt photon 
hadroproduction $p\bar p \to \gamma\,X$ at Tevatron come from both the
D$\oslash$ and CDF collaborations. The D$\oslash$~[1, 2] data 
were obtained in the central and forward pseudo-rapidity
regions for two different center-of-mass energies, namely 
$\sqrt s = 630$ GeV and $\sqrt s = 1800$ GeV.
The central pseudo-rapidity region is defined by the requirement
$|\eta| < 0.9$, and the forward one is 
defined by $1.6 < |\eta| < 2.5$.
The more recent CDF data~[3, 4] refer to the same central kinematical region
$|\eta| < 0.9$  
for both beam energies $\sqrt s = 630$ GeV and $\sqrt s = 1800$ GeV.
As usually, the photon pseudo-rapidity $\eta$ is defined as 
$\eta = - \ln \tan (\theta/2)$, where $\theta$ is the polar 
angle of the prompt photon with respect to the proton beam.
Also the CDF collaboration has presented a measurement~[5] 
of the prompt photon cross section at $\sqrt s = 1800$ GeV which
is based on events where the photon converts
into an electron-positron pair in the material of inner detector,
resulting in a two track event signature ("conversion" data).
These data refer only to the central kinematical region. 
Actually, there are two different datasets, 
which were used in the CDF measurement with conversions,
namely 8 GeV electron data and 23 GeV photon 
data\footnote{See Ref.~[5] for more details.}.
Last available experimental data has been presented
by the D$\oslash$ collaboration very recently~[6].
These data extend previous measurements~[1--5] to significantly 
higher values of photon $p_T$ (namely to $p_T \sim 300$~GeV at 
$\sqrt s = 1960$~GeV).

The results of our calculations are shown in Figs.~3 --- 18. 
So, Figs.~3 and~4 confront the double differential cross sections 
$d\sigma/dE_T d \eta$ calculated at $\sqrt s = 630$ GeV 
in different kinematical regions with the D$\oslash$~[2] and CDF~[3] data.
The solid and dottes lines are obtained by using the CCFM and KMR 
unintegrated gluon densities with the default factorization and 
renormalization scale. The upper and lower dashed 
lines correspond to the scale variations in CCFM gluon 
as it was described above. The data/theory ratio is also shown in Figs.~5 and~6.
One can see that
our predictions agree well with the experimental data within the 
scale uncertainties. However, the results of calculation with the 
KMR gluon at the default scale tend to underestimate the data in 
the central kinematical region and agree with the D$\oslash$ 
data in the forward $\eta$ region. The CCFM gluon density
gives a perfect description of the data in both kinematical regions.
Main difference in predictions of the KMR and CCFM gluon
densities concentrated in the central pseudo-rapidity region at low $p_T$.
This region corresponds to the low $x$ values where effects of 
the gluon evolution play the dominant role. 
It is clear from the Figs.~7 and~8 where gluon and valence 
quark contributions are shown separately. In the central
pseudo-rapidity region the gluon-gluon fusion dominates 
up to $p_T \sim 20$~GeV approximately whereas in the
forward pseudo-rapidity region the valence quark contribution
should be taken into account everywhere.
In these plots we have used the CCFM gluon density 
for illustration. 
Note that the KMR gluon 
give description of the data which is rather similar to collinear 
NLO QCD calculations~[17]: 
the results of measurement are higher than 
the NLO prediction at low $E_T$ in the central $\eta$ range 
but agree at all $E_T$ in the forward pseudo-rapidity region.

It is important that KMR predictions based 
on the $g^* g^* \to q\bar q \gamma$ 
off-shell matrix element practically coincide with 
the previous ones~[32] which were based on the quark-gluon fusion 
and quark-antiquark annihilation. This fact gives an additional
check of our calculations and the self-consistency of the 
whole KMR scheme.

The double differential cross sections $d\sigma/dE_T d \eta$ 
compared with the experimental data at $\sqrt s = 1800$ GeV 
in different pseudo-rapidity regions are shown in 
Figs.~9 --- 11. The data/theory ratio is depicted in Figs.~12 and~13.
The gluon and valence quark contributions to the 
cross section at this energy are shown separately in
Fig.~14 and~15. We find that our predictions with CCFM gluon
density reasonable agree with the D$\oslash$~[1] and CDF~[3--5] 
data both in normalization and shape. However, the using of the KMR-evolved
gluon density results to underestimation of the measurement by a factor about~2 in
the central $\eta$ region. The similar situation is 
observed at $\sqrt s = 1960$~GeV (see Fig.~16).
This is due to different small-$x$ 
behaviour of the unintegrated gluon densities, since valence
quark contribution in this region is about~7\% only, as it was 
demonstrated in Fig.~14. In forward pseudo-rapidity region 
the results obtained with the both gluon densities under consideration
agree well with the experimental data. 
So one can conclude that in the central 
pseudo-rapidity region at low $p_T$ the observed cross section
is a strongly sensitive to the unintegrated gluon density used.
Therefore available experimental data can be applied in future 
to better constraint of the unintegrated gluon distribution. Specially it will 
useful in the case of the CCFM evolution (which is valid for 
both small and large values of $x$) since measurements~[1--6] on the prompt photon 
production refers to whole $x$ range.
One can see that the gluon-gluon contribution at $\sqrt s = 1800$~GeV 
dominates up to $p_T \sim 50$~GeV in the central $\eta$ region
and up to $p_T \sim 20$~GeV in the forward one, where 
$x$ is small still. In general, the contribution from valence quarks is more
significant in the forward rapidity region.
This situation is the similar to one at $\sqrt s = 630$~GeV.

Concerning the collinear approximation of QCD,
it was shown~[6] that results from NLO calculations
agree with the recent high $p_T$ measurement within 
uncertainties. However, at moderate and small $p_T$ region
(which corresponds to small values of $x$) the shape of the 
measured cross sections at $\sqrt s = 1800$~GeV 
is generally steeper than that of the NLO predictions.
It was claimed~[3, 4] that this shape difference 
is difficult to explain simply by changing the 
renormalization/factorization scales in the collinear calculation, 
or the set of parton distribution functions.

Now we can try to extrapolate our theoretical predictions
to LHC energies. We perform the calculation for both 
central and forward photon pseudo-rapidities $\eta$.
As a representative example,
we will define the central and forward kinematical regions 
by the requirements $|\eta| < 2.5$ and 
$2.5 < |\eta| < 4$, respectively.
The transverse energy $E_T$
distributions of inclusive prompt photon production at 
$\sqrt s = 14$ TeV are shown in Figs.~17 and~18. 
The formulas~(15) were used in these predictions.
These Figs show the size of 
theoretical uncertainties connected with
the unintegrated gluon densities. 
It is worth mentioning that the extrapolation of the available parton 
distribution to the region of lower $x$ is a special 
problem at the LHC energies. In particular, one of the problem
is connected with the correct treatment of saturation effects
in small $x$ region\footnote{See, for example, Ref.~[35]}. 
Therefore additional work needs to be done
until these uncertainties will be reduced.

\section{Conclusions} \indent 

We have tried a new theoretical approach to the production of prompt 
photons in hadronic collisions at high energies. 
Our approach is based on the $k_t$-factorization scheme, which, 
unlike many early calculations~[21, 22, 25], provides solid theoretical grounds 
for adequately taking into account the effects of initial parton 
momentum. The central part of our consideration is the off-shell gluon-gluon 
fusion subprocess $g^*g^*{\to}q\bar{q}\gamma$. The corresponding off-shell 
matrix elements have been calculated for the first time. At the price of 
considering the $2{\to}3$ rather than $2{\to}2$ matrix elements, we have reduced 
the problem of unknown unintegrated quark distributions to the 
problem of gluon distributions. 
This way enables us with making comparisons between the different parton 
evolution schemes and parametrizations of parton densities, in 
contrast with previous calculations~[18, 32] where such a comparison was not 
possible (for the lack of unintegrated quark distributions except KMR). 

Since the gluons are only responsible for the appearance of sea, but 
not valence quarks, the contribution from the valence quarks has been 
calculated separately. Having in mind that the valence quarks are only 
important at large $x$, where the traditional DGLAP evolution is accurate and 
reliable, we have calculated this contribution within the usual collinear 
scheme based on $2\to 2$ partonic subprocesses and on-shell parton densities. 
             
We have calculated the total and differential cross sections 
and have made comparisons to the recent D$\oslash$ and CDF 
experimental data. In the numerical analysis we have used 
the unintegrated gluon densities which are obtained from the full 
CCFM equation as well as from the KMR prescription. 
We have found that in the central pseudo-rapidity region at low $p_T$ the 
observed cross section is strongly sensitive to the unintegrated 
gluon density. In this kinematic region the cross sections
evaluated with CCFM and KMR gluon distributions differ from 
each other by a factor of about 2. 
This is due to the fact that low-$p_T$ measurements in the 
central pseudo-rapidity region probe the small-$x$ region,
where the shapes of the considered gluon densities are
different.
In the forward pseudo-rapidity region the
predictions from different gluon densities practically coincide.
We have demonstrated that the available experimental 
data can be used to constrain the unintegrated 
gluon distributions. Especially, it may be useful in the case of  
CCFM evolution (which is valid for both small and large values of $x$) 
since the recent D$\oslash$ and CDF measurements cover the whole $x$ range.

Our results based on Kimber-Martin-Ryskin~[31] gluon density and off-shell 
gluon-gluon fusion matrix element $g^*g^*{\to}q\bar{q}\gamma$ agree with the previous 
results~[32] based on the quark-gluon fusion $qg{\to}q\gamma$ and quark-antiquark 
annihilation $q\bar{q}{\to}g\gamma$ matrix elements and KMR~[31] 
off-shell quark and gluon distributions. This can be regarded as an additional 
proof of the consistency of the proposed method. 

\section{Acknowledgements} \indent 

We thank H.~Jung for encouraging interest, very helpful discussions
and for providing the CCFM code for 
unintegrated gluon distributions. 
The authors are very grateful to P.F.~Ermolov for the support and
DESY Directorate for the support in the 
framework of Moscow --- DESY project on Monte-Carlo
implementation for HERA --- LHC.
A.V.L. was supported in part by the grants of the president of 
Russian Federation (MK-9820.2006.2) and Helmholtz --- Russia
Joint Research Group.
Also this research was supported by the 
FASI of Russian Federation (grant NS-8122.2006.2).

\newpage

\begin{figure}
\begin{center}
\epsfig{figure=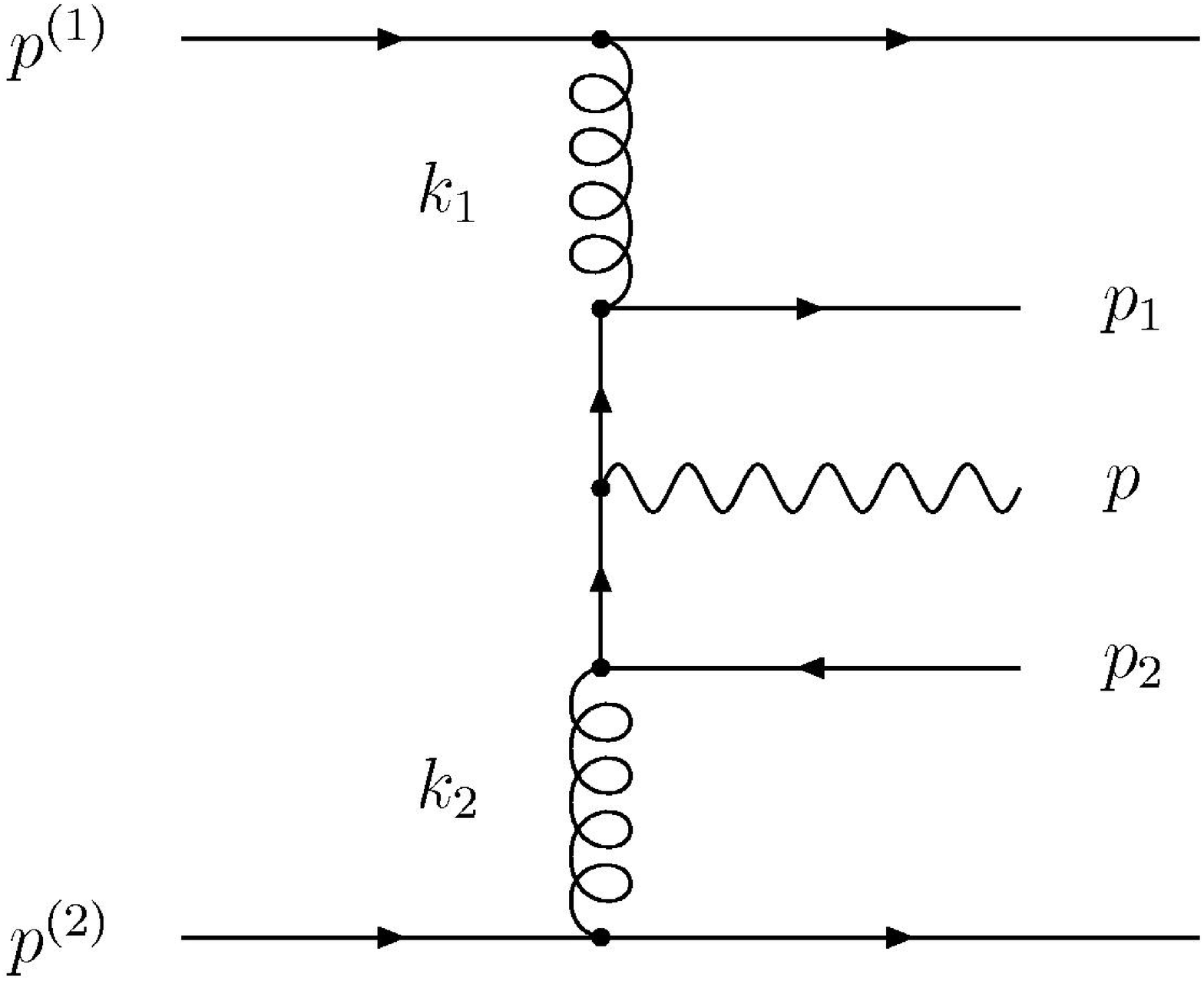, width = 8cm}
\caption{Kinematics of the $g^* g^* \to q \bar q \gamma$ process.}
\label{fig1}
\end{center}
\end{figure}

\newpage

\begin{figure}
\begin{center}
\epsfig{figure=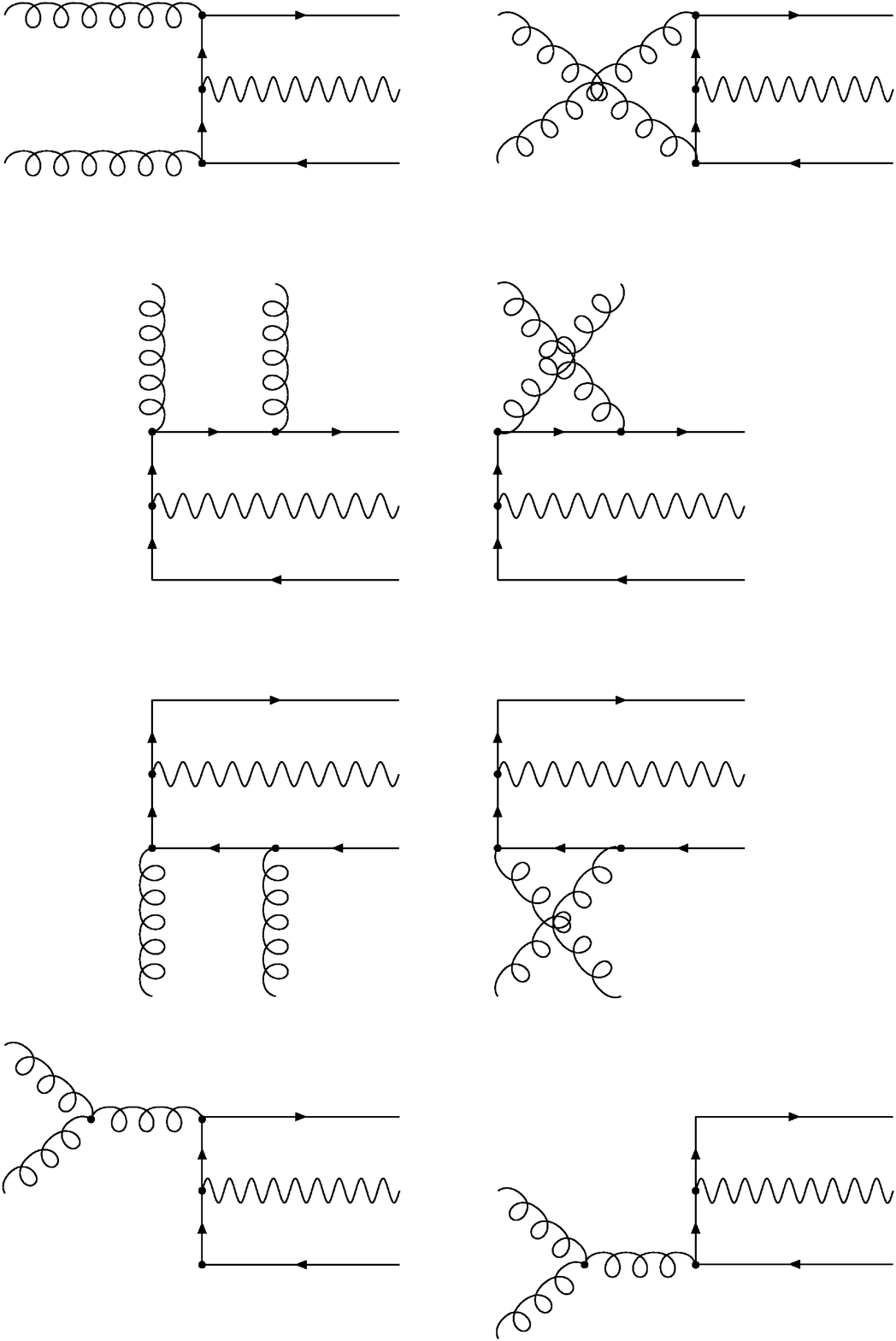, width = 14cm}
\caption{Feynman diagrams which describe the partonic
subprocess $g^* g^* \to q \bar q \gamma$ at the leading order in $\alpha_s$ and $\alpha$.}
\label{fig2}
\end{center}
\end{figure}

\newpage

\begin{figure}
\begin{center}
\epsfig{figure=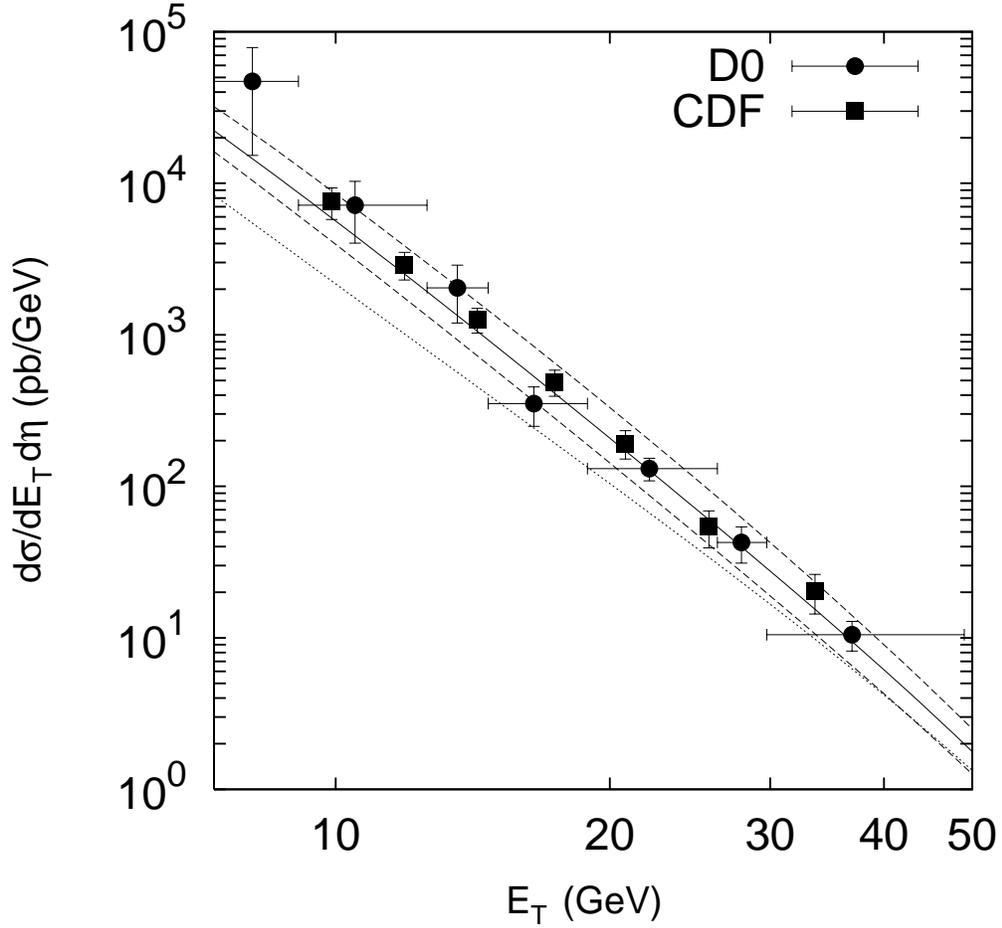, width = 18cm}
\caption{The double differential cross section $d\sigma/d E_T d\eta$
for inclusive prompt photon hadroproduction 
at $|\eta| < 0.9$ and $\sqrt s = 630$ GeV. The solid line corresponds to the CCFM 
gluon density with the default scale $\mu = E_T$, whereas upper and lower dashed lines
correspond to the usual scale variation in the CCFM distribution, respectively. 
The dotted line corresponds to the KMR unintegrated gluon density.
The experimental data are from D$\oslash$~[2] and CDF~[3].}
\end{center}
\label{fig3}
\end{figure}

\newpage

\begin{figure}
\begin{center}
\epsfig{figure=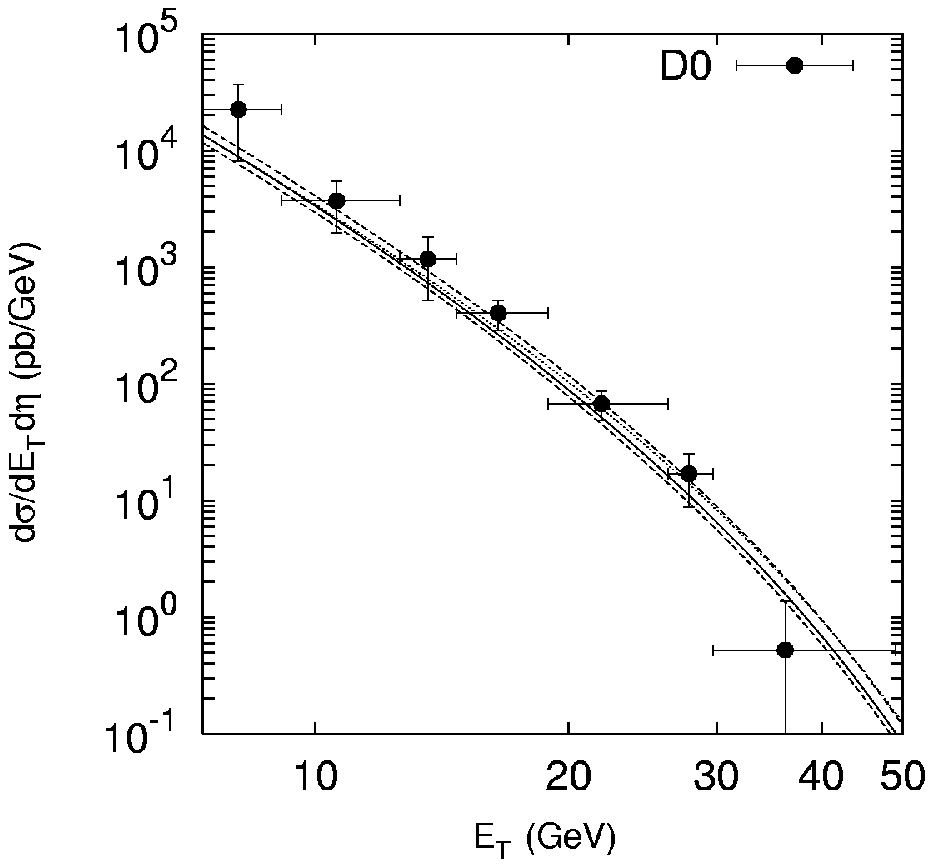, width = 18cm}
\caption{The double differential cross section $d\sigma/d E_T d\eta$
for inclusive prompt photon hadroproduction 
at $1.6 < |\eta| < 2.5$ and $\sqrt s = 630$ GeV. 
Notations of all curves are the same as in Figure~3.
The experimental data are from D$\oslash$~[2].}
\end{center}
\label{fig4}
\end{figure}

\newpage

\begin{figure}
\begin{center}
\epsfig{figure=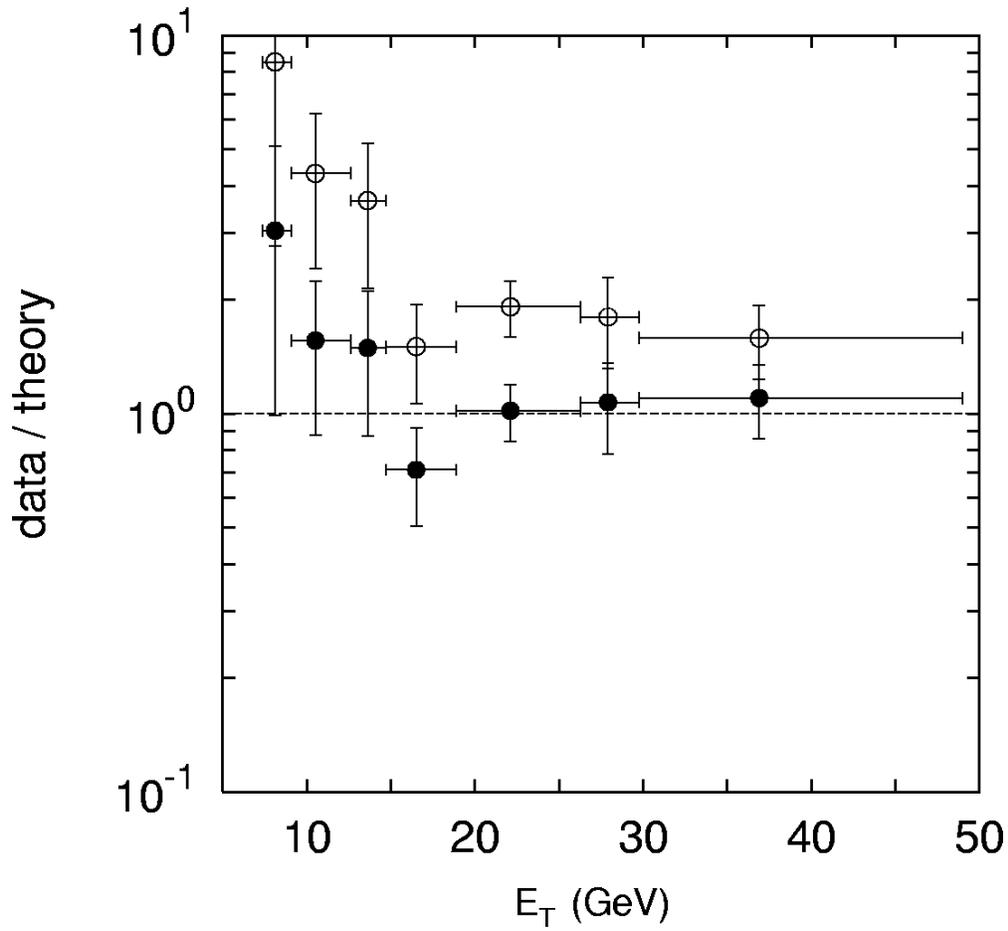, width = 18cm}
\caption{The data/theory ratio for inclusive prompt photon hadroproduction 
calculated at $|\eta| < 0.9$ and $\sqrt s = 630$ GeV. Black and open circles
correspond to the CCFM and KMR gluon densities, respectively.
The experimental data are from D$\oslash$~[2].}
\end{center}
\label{fig5}
\end{figure}

\newpage

\begin{figure}
\begin{center}
\epsfig{figure=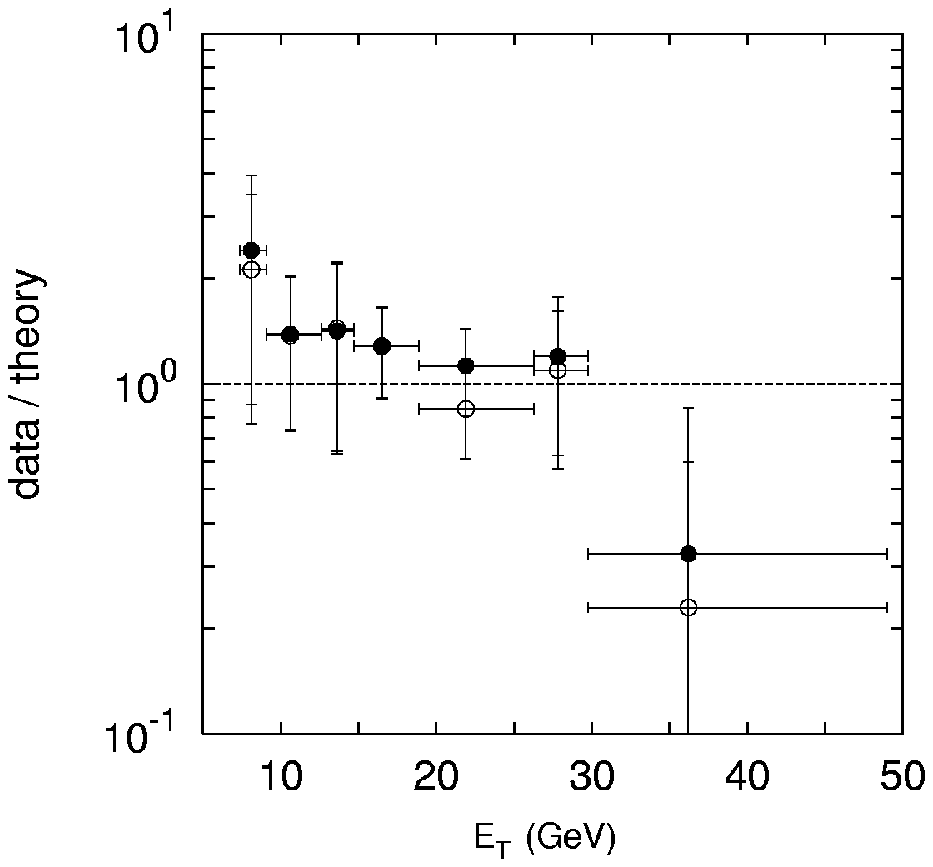, width = 18cm}
\caption{The data/theory ratio for inclusive prompt photon hadroproduction 
calculated at $1.6 < |\eta| < 2.5$ and $\sqrt s = 630$ GeV. 
Notations are the same as in Figure~5.
The experimental data are from D$\oslash$~[2].}
\end{center}
\label{fig6}
\end{figure}

\newpage

\begin{figure}
\begin{center}
\epsfig{figure=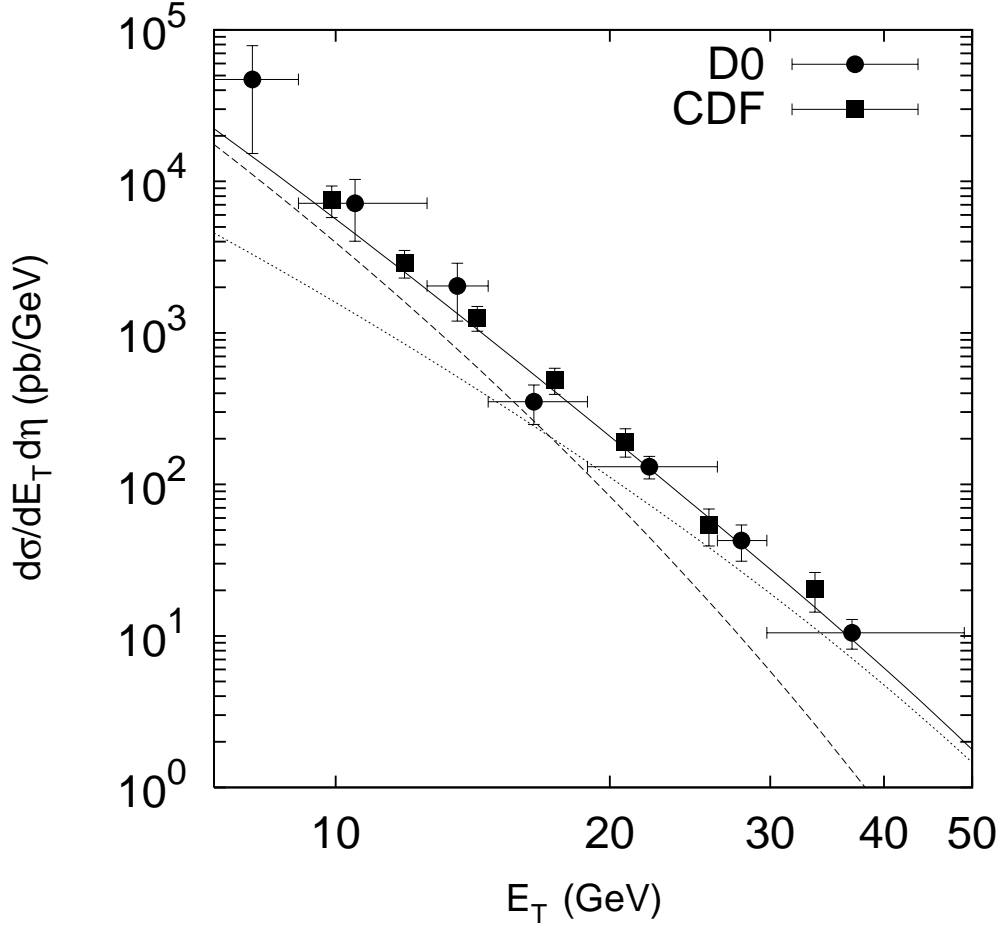, width = 18cm}
\caption{The double differential cross section $d\sigma/d E_T d\eta$
for inclusive prompt photon hadroproduction 
at $|\eta| < 0.9$ and $\sqrt s = 630$ GeV. The dashed and dotted lines 
represent the gluon and valence quark contributions to the prompt
photon cross section, respectively. The solid line corresponds to 
the sum of these contributions. We have use here the CCFM gluon
densities. The experimental data are from D$\oslash$~[2] and CDF~[3].}
\end{center}
\label{fig7}
\end{figure}

\newpage

\begin{figure}
\begin{center}
\epsfig{figure=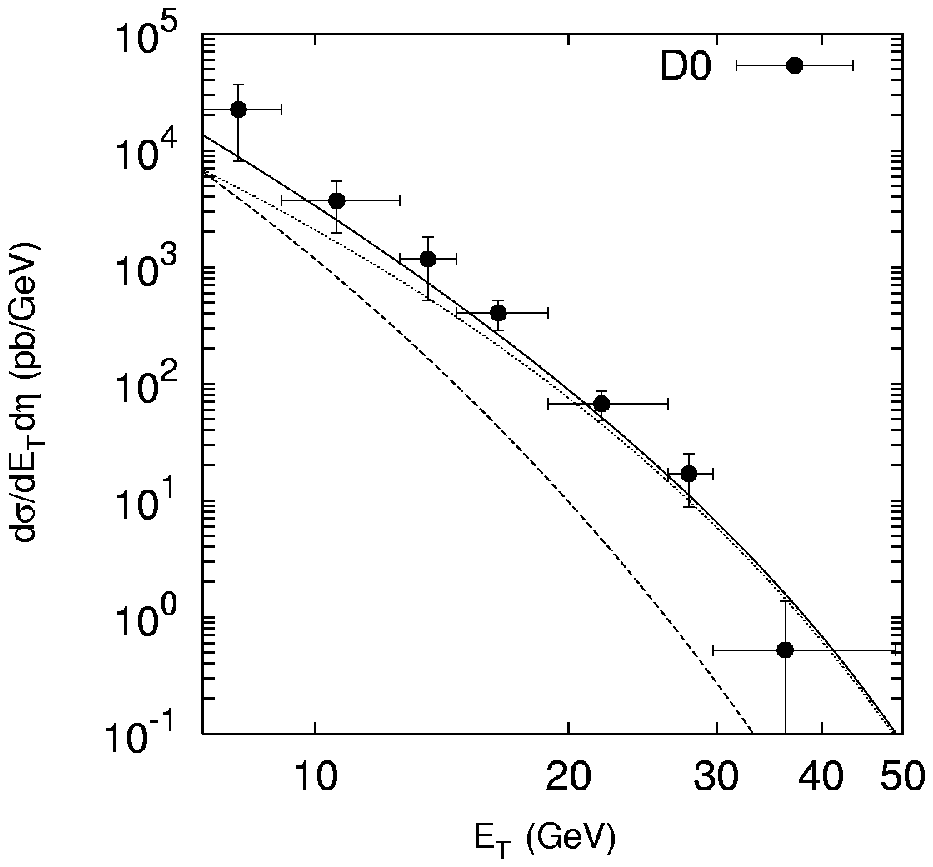, width = 18cm}
\caption{The double differential cross section $d\sigma/d E_T d\eta$
for inclusive prompt photon hadroproduction 
at $1.6 < |\eta| < 2.5$ and $\sqrt s = 630$~GeV. 
Notations of all curves are the same as in Figure~7.
The experimental data are from D$\oslash$~[2].}
\end{center}
\label{fig8}
\end{figure}

\newpage

\begin{figure}
\begin{center}
\epsfig{figure=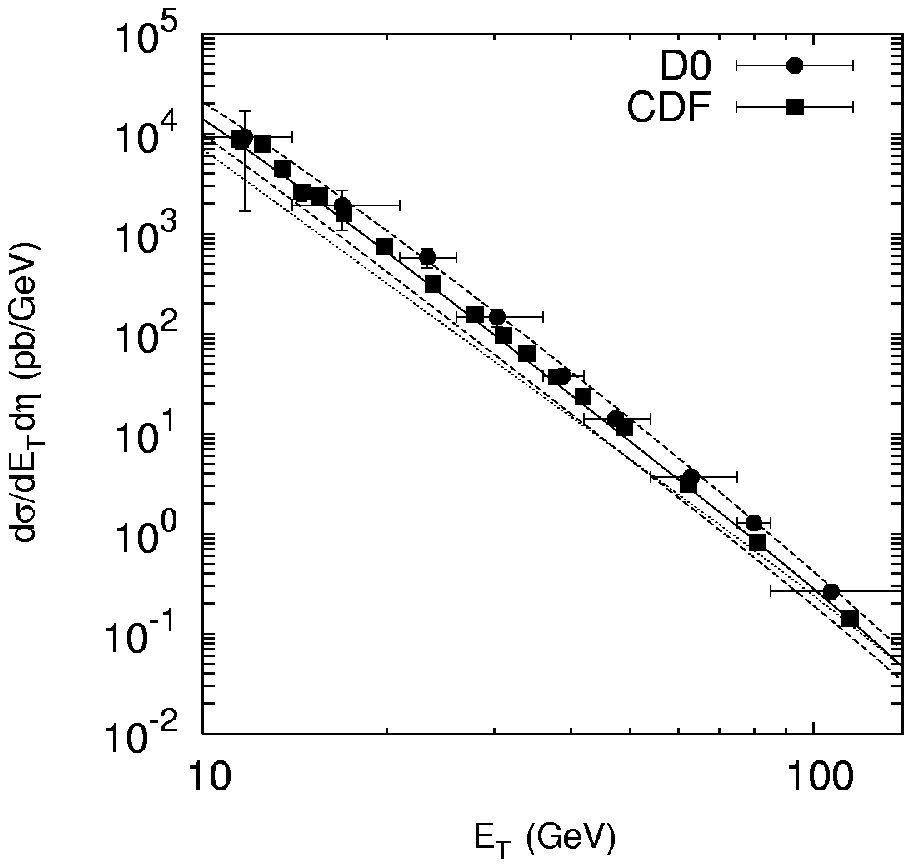, width = 18cm}
\caption{The double differential cross section $d\sigma/d E_T d\eta$
for inclusive prompt photon hadroproduction 
at $|\eta| < 0.9$ and $\sqrt s = 1800$~GeV. 
Notations of all curves are the same as in Figure~3.
The experimental data are from D$\oslash$~[1] and CDF~[3, 4].}
\end{center}
\label{fig9}
\end{figure}

\newpage

\begin{figure}
\begin{center}
\epsfig{figure=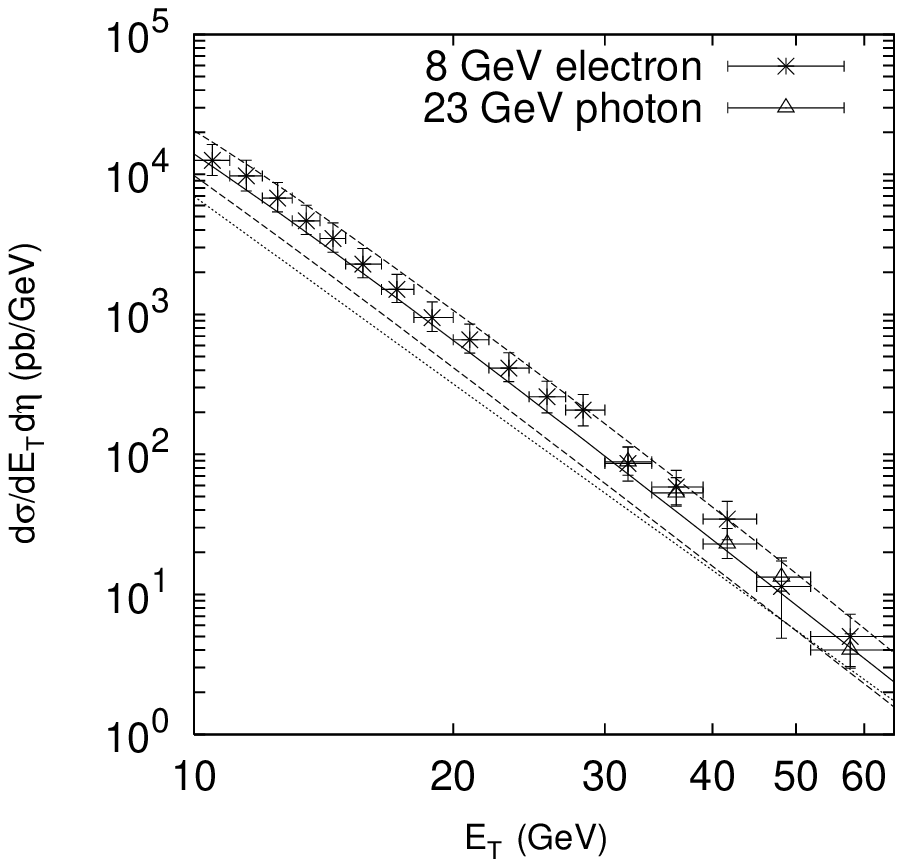, width = 18cm}
\caption{The double differential cross section $d\sigma/d E_T d\eta$
for inclusive prompt photon hadroproduction 
at $|\eta| < 0.9$ and $\sqrt s = 1800$~GeV. 
Notations of all curves are the same as in Figure~3.
The experimental data are from CDF~[5].}
\end{center}
\label{fig10}
\end{figure}

\newpage

\begin{figure}
\begin{center}
\epsfig{figure=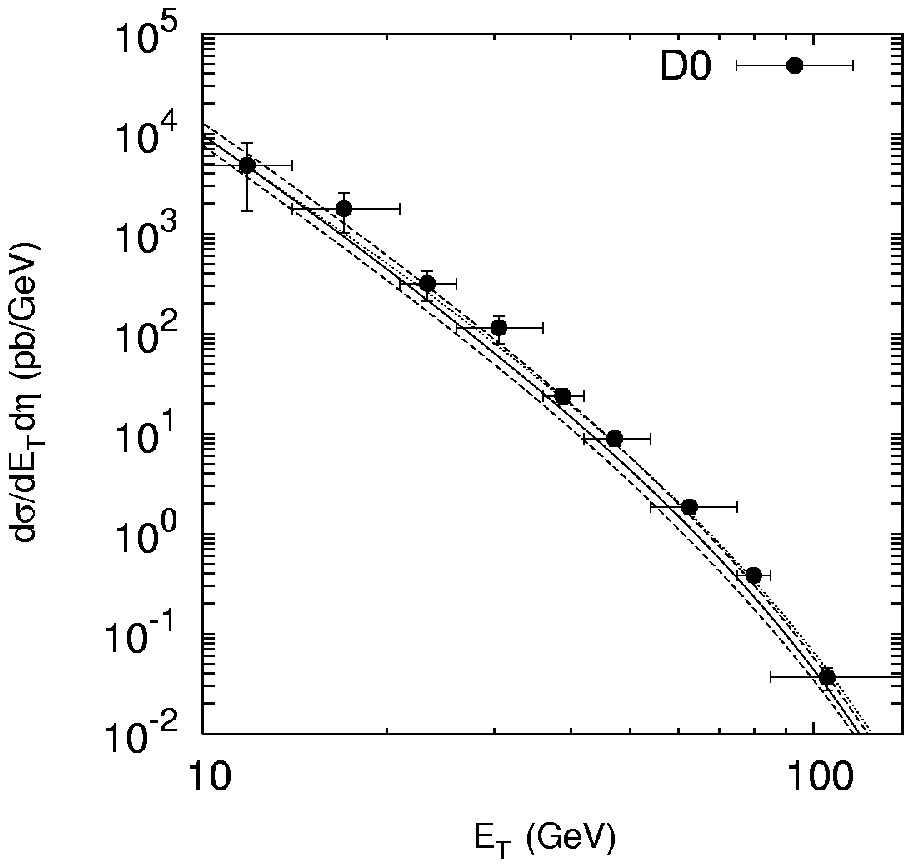, width = 18cm}
\caption{The double differential cross section $d\sigma/d E_T d\eta$
for inclusive prompt photon hadroproduction 
at $1.6 < |\eta| < 2.5$ and $\sqrt s = 1800$~GeV. 
Notations of all curves are the same as in Figure~3.
The experimental data are from D$\oslash$~[1].}
\end{center}
\label{fig11}
\end{figure}

\newpage

\begin{figure}
\begin{center}
\epsfig{figure=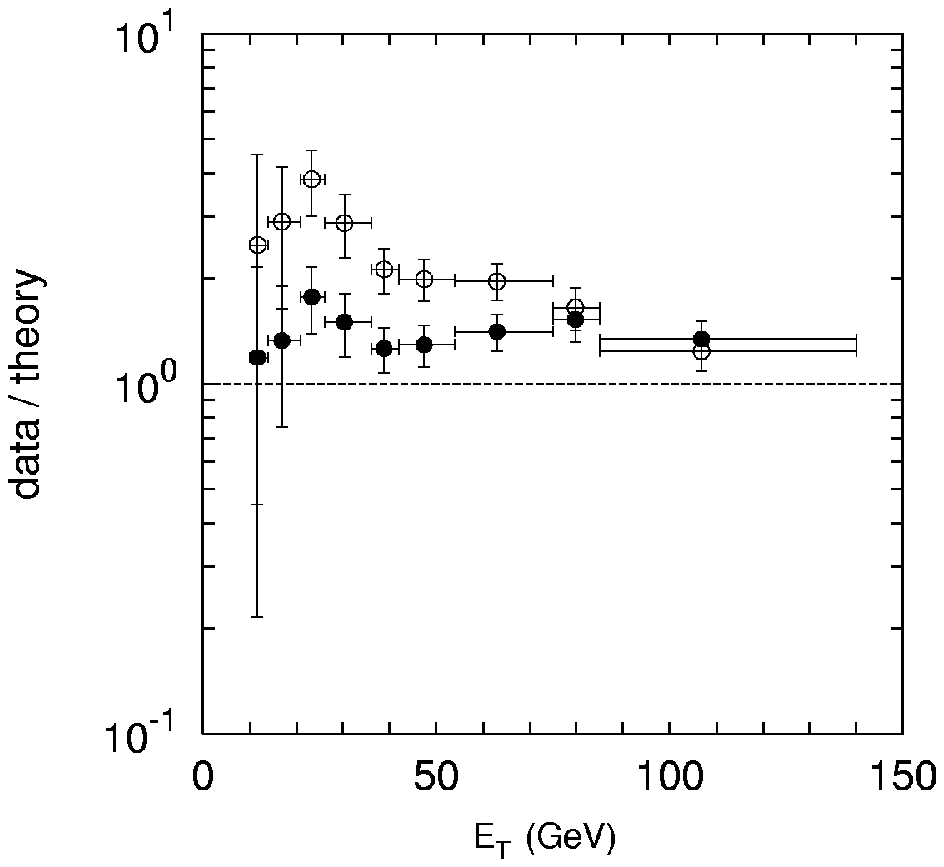, width = 18cm}
\caption{The data/theory ratio for inclusive prompt photon hadroproduction 
calculated at $|\eta| < 0.9$ and $\sqrt s = 1800$ GeV. 
Notations are the same as in Figure~5.
The experimental data are from D$\oslash$~[1].}
\end{center}
\label{fig12}
\end{figure}

\newpage

\begin{figure}
\begin{center}
\epsfig{figure=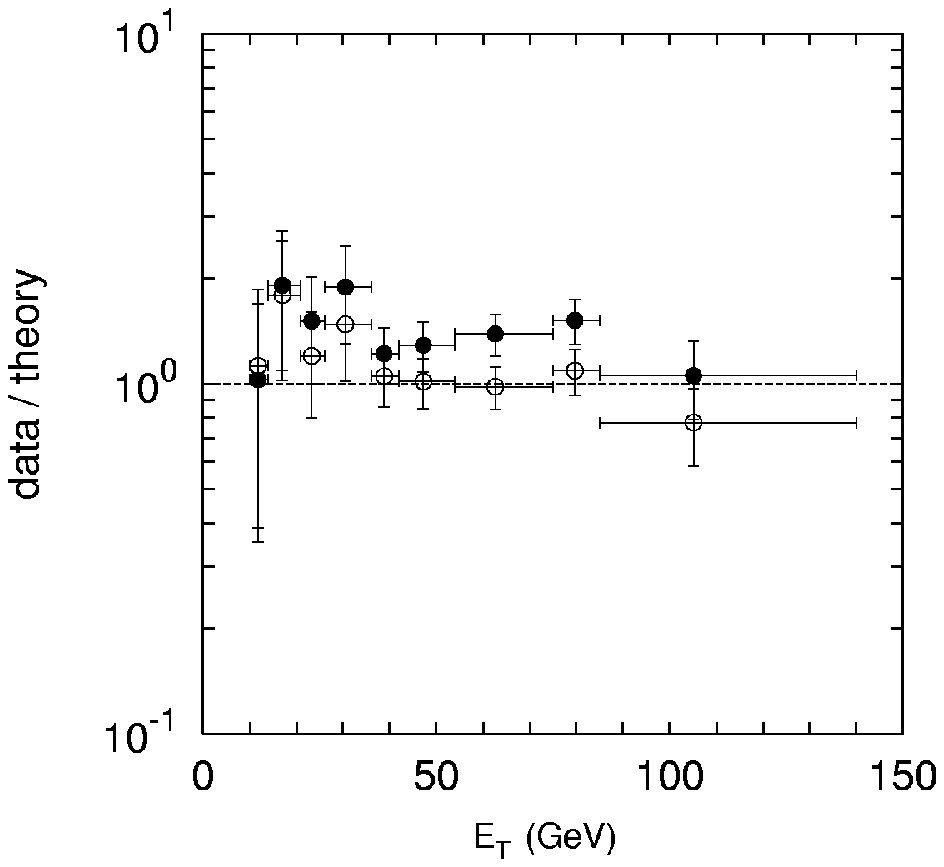, width = 18cm}
\caption{The data/theory ratio for inclusive prompt photon hadroproduction 
calculated at $1.6 < |\eta| < 2.5$ and $\sqrt s = 1800$ GeV. 
Notations are the same as in Figure~5.
The experimental data are from D$\oslash$~[1].}
\end{center}
\label{fig13}
\end{figure}

\newpage

\begin{figure}
\begin{center}
\epsfig{figure=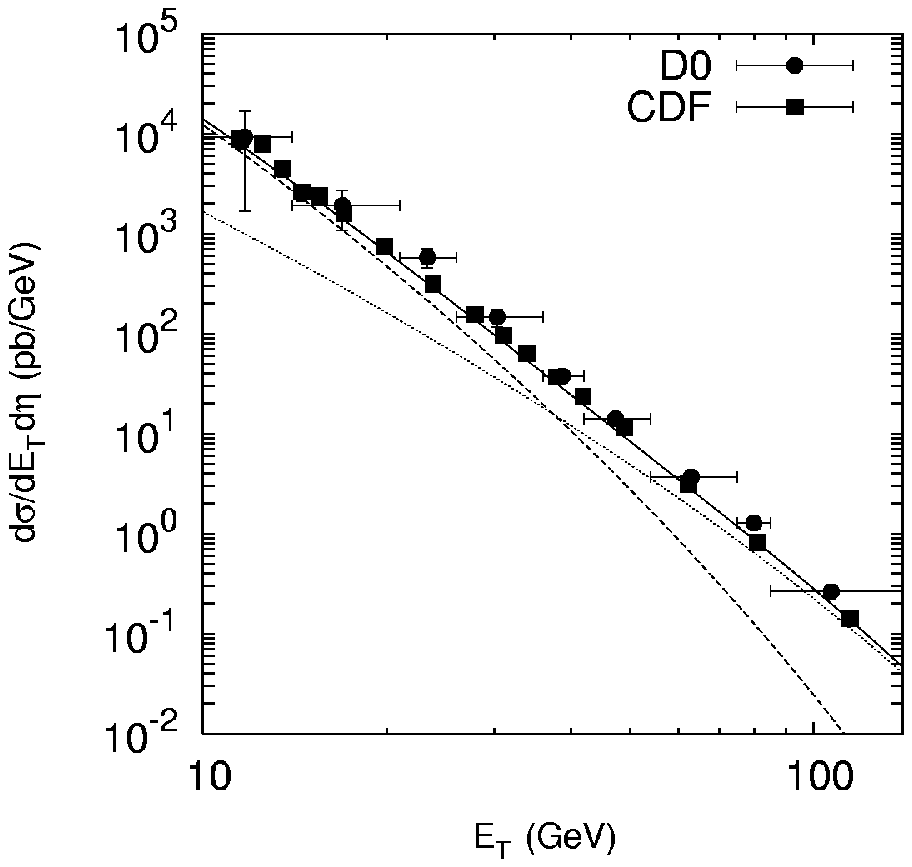, width = 18cm}
\caption{The double differential cross section $d\sigma/d E_T d\eta$
for inclusive prompt photon hadroproduction 
at $|\eta| < 0.9$ and $\sqrt s = 1800$~GeV. 
Notations of all curves are the same as in Figure~7.
The experimental data are from D$\oslash$~[1] and CDF~[3, 4].}
\end{center}
\label{fig14}
\end{figure}

\newpage

\begin{figure}
\begin{center}
\epsfig{figure=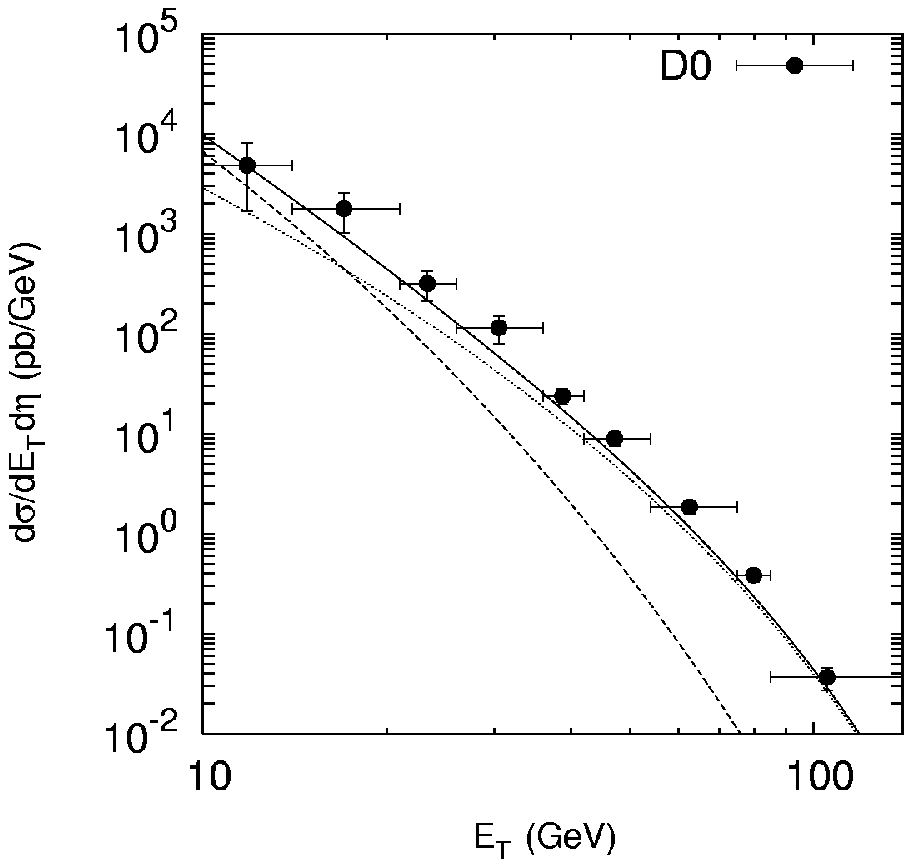, width = 18cm}
\caption{The double differential cross section $d\sigma/d E_T d\eta$
for inclusive prompt photon hadroproduction 
at $1.6 < |\eta| < 2.5$ and $\sqrt s = 1800$~GeV. 
Notations of all curves are the same as in Figure~7.
The experimental data are from D$\oslash$~[1].}
\end{center}
\label{fig15}
\end{figure}

\newpage

\begin{figure}
\begin{center}
\epsfig{figure=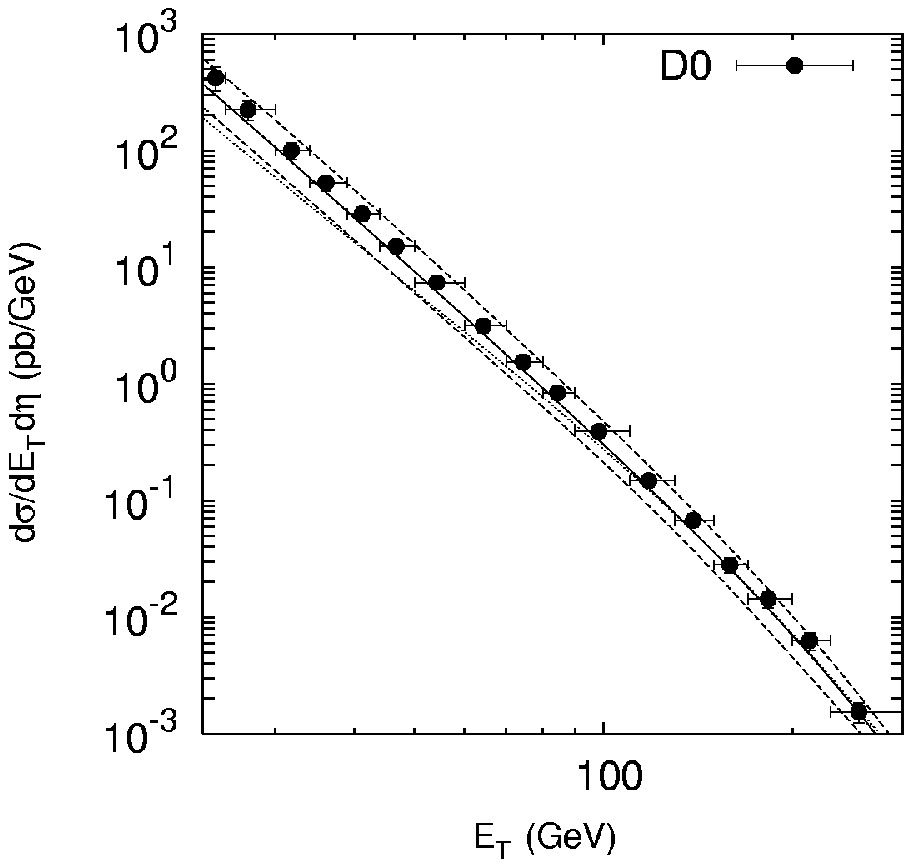, width = 18cm}
\caption{The double differential cross section $d\sigma/d E_T d\eta$
for inclusive prompt photon hadroproduction 
at $|\eta| < 0.9$ and $\sqrt s = 1960$~GeV. 
Notations of all curves are the same as in Figure~3.
The experimental data are from D$\oslash$~[6].}
\end{center}
\label{fig16}
\end{figure}

\newpage

\begin{figure}
\begin{center}
\epsfig{figure=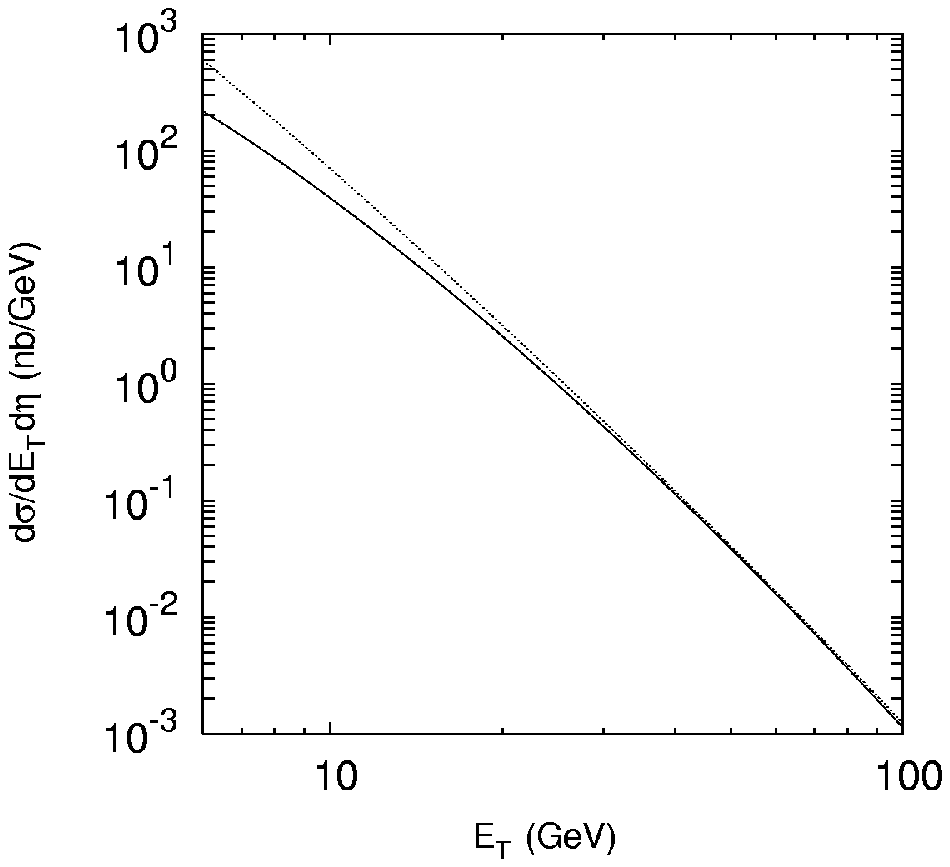, width = 18cm}
\caption{The transverse enery $E_T$ distribution of inclusive prompt 
photon hadroproduction calculated at $|\eta| < 2.5$ and $\sqrt s = 14$~TeV. 
Notations of all curves are the same as in Figure~3.}
\end{center}
\label{fig17}
\end{figure}

\newpage

\begin{figure}
\begin{center}
\epsfig{figure=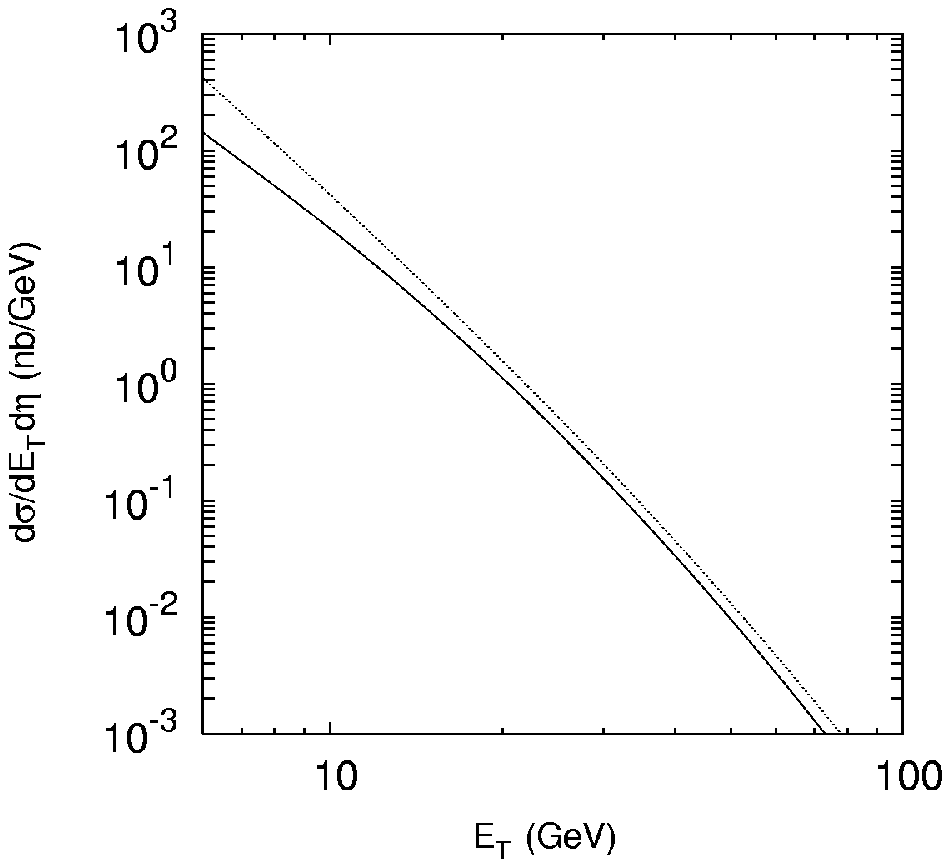, width = 18cm}
\caption{The transverse enery $E_T$ distribution of inclusive prompt 
photon hadroproduction calculated at $2.5 < |\eta| < 4$  
and $\sqrt s = 14$~TeV. 
Notations of all curves are the same as in Figure~3.}
\end{center}
\label{fig18}
\end{figure}

\end{document}